\documentclass[aps,twocolumn,showpacs,floatfix]{revtex4}

\usepackage{dcolumn}
\usepackage{graphicx}
\usepackage{amsfonts}
\usepackage{amsmath,bm,amssymb}

\begin{document}

\title{Doping Driven ($\pi ,0$) Nesting and Magnetic Properties of Fe$_{1+x}$%
Te Superconductors}
\author{Myung Joon Han and Sergey Y. Savrasov}
\affiliation{Department of Physics, University of California, Davis, California 95616, USA}
\date{\today }

\begin{abstract}
To understand newly discovered superconductivity in Fe--based systems, we
investigate electronic structure and magnetic properties of Fe$_{1+x}$Te
using first--principles density functional calculations. While the undoped
FeTe has the same Fermi surface nested at ($\pi ,\pi $) as in Fe arsenides,
doping by $\sim 0.5$ electrons reveals a novel square--type Fermi surface
showing a strong ($\pi ,0$) nesting and leading to a different magnetic
structure. Our result strongly supports the same mechanism of
superconductivity in chalcogenides as in the arsenides, reconciling theory
to existing experiments. Calculated magnetic interactions are found to be
critically dependent on doping and notably different from the arsenides.
\end{abstract}

\maketitle

Shortly after the discovery of a novel high temperature superconductor,
LaFeAsO$_{1-x}$F$_{x}$ with T$_{c}\sim $27 K \cite{JACS}, many different
types of iron--based superconductors have been reported. Now the highest T$%
_{c}$ reaches up to $\sim $55 K \cite{Sm-55K}, and there are four different
structural classes: so--called 122--\cite{Rotter-122}, 111--\cite{Tapp-111},
and 11--type\cite{FeSe} structures besides the original 1111 type. Although
tremendous research activities devoted to this field over the year have shed
light on their intriguing physical properties, our understanding of
superconductivity here and its interplay with magnetism is still far from
being complete. One of the most important properties which was found in
these systems is the Fermi surface nesting whose nesting vector corresponds
to the antiferromagnetic (AFM) ordering vector of the undoped magnetic phase 
\cite{Singh-review,Mazin-review,epl,Mazin}. Density Functional Theory (DFT)
calculations show that all of the four classes of these materials share this
common feature in their electronic structure \cite{Mazin,Singh,Subedi},
which strongly suggests the superconductivity is  exotic and is mediated by
spin fluctuations \cite{Mazin-review}.

Along this line, one of the most interesting questions arises in the
11--type Fe chalcogenide family: Fe(S,Se,Te) \cite%
{FeSe,FeSe-27K,FeSeTe-1,FeSeTe-2,FeSeTe-Incomm,Mizuguchi-FeTe,Mizuguchi-S,Mizuguchi-S-Se-Te}%
. In spite of their same crystal structure represented by the 2--dimensional
Fe square lattice and the same Fermi surface nesting predicted by DFT
calculations \cite{Subedi}, Fe chalcogenide superconductors exhibit notable
differences from the arsenides. The magnetic structure found in their parent
compound, Fe$_{1+x}$Te, strongly tackles the spin fluctuation theory because
the experimentally observed magnetic ordering is fairly different from that
of parent arsenide compounds. Although DFT calculations predict the same
Fermi surface topology, a recent neutron experiment \cite{Li} shows that, at
a small $x\sim 0.068$, Fe$_{1+x}$Te has a rotated and double stripe AFM
order. Importantly, this magnetic structure cannot be matched with the ($\pi
,\pi $) nesting as found in all the arsenide materials and predicted by
previous calculations, but requires ($\pi ,0$) nesting which has never been
reported. This `missing nesting' remains\ as a puzzle in the study of
Fe--based superconductors. Therefore it is not a big surprise that some
papers speculate about a different superconducting mechanism for Fe
chalcogenides from the arsenides, and the reinvestigation of the electronic
structure and magnetic properties for Fe$_{1+x}$Te is of crucial importance 
\cite{FeSeTe-Incomm,Xia,McQueen,Fang}.

To address these issues, we study electronic structure and magnetic
interactions of Fe$_{1+x}$Te using first--principles DFT calculations. Since
it is noted experimentally \cite{Li} that FeTe has always some amount of
excess Fe atoms we perform doping dependent calculations to understand their
effect on the electronic and magnetic properties. Our results show that Fe$%
_{1+x}$Te has a different Fermi surface topology as a function of doping,
and eventually a novel ($\pi ,0$) nesting appears at a doping level of $%
\delta \sim 0.5$ electrons while the ($\pi ,\pi $) nesting is largely
suppressed. This ($\pi ,0$) nesting exactly matches with the double--stripe
AFM order found in neutron experiments. Our result strongly supports the
spin fluctuation mediated superconductivity for Fe chalcogenides,
reconciling theory to existing experiments which showed significant
differences between pnictides and chalcogenides. It is also found that
magnetic interactions depend on doping, and, at the same level of doping $%
\delta \sim 0.5$, the calculated exchange couplings become consistent with
the double stripe phase. While the second nearest neighbor AFM coupling ($%
J_{2a}$) is strongest, the first ($J_{1b}$) and second neighbor
ferromagnetic (FM) interactions ($J_{2b}$) are also significant, which is
different from arsenides.

\begin{figure}[t]
\centering \includegraphics[width=8.3cm]{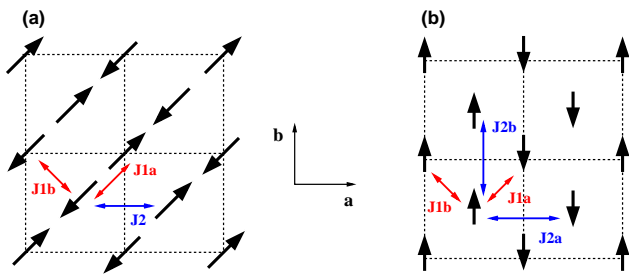}
\caption{(Color online) The schematic picture of spin structure for (a) iron
arsenides parent materials and (b) FeTe. The dotted squares correspond to
the non--magnetic unit cell and the arrows represent the spin directions.
The $J_{1a}$, $J_{1b}$, $J_{2a}$, and $J_{2b}$ refer to the nearest neighbor
AFM, nearest neighbor FM, next nearest neighbor AFM, and next nearest
neighbor FM exchange interaction, respectively. }
\label{schem}
\end{figure}

We performed the first--principles DFT calculations within local density
approximation (LDA) for the exchange--correlation energy functionals \cite%
{LDA}. The full potential linearized--muffin--tin--orbital (LMTO) method has
been used \cite{Savrasov-1996}. To estimate the exchange interaction
strengths between Fe moments, we performed a linear response calculation 
\cite{J,J2,Wan}, which has been successfully applied earlier to $3d$
transition--metal oxides, $5f$ metallic alloys \cite{Wan,Han}, and the other
Fe arsenides \cite{Han2}. To simulate electron doping a rigid--band
approximation is utilized. The calculation of band dispersions, Fermi
surfaces, and Stoner response functions have been done with the
non--magnetic unit cell (dotted squares in Fig.~\ref{schem}) and using
experimental lattice parameters as in the previous study \cite{Subedi}. To
estimate the exchange constants we performed a spin--polarized calculation
with an enlarged unit cell containing four Fe atoms, and the experimental $%
z(Te)$ was used which well reproduces the observed moment and is consistent
with our previous study of Fe arsenides \cite{Han2}.

Fig.~\ref{schem} summarizes spin structures found in the parent materials of
arsenide superconductors (Fig.~\ref{schem}(a)) and Fe$_{1+x}$Te for small $x$
(Fig.~\ref{schem}(b)) \cite{Li}. In Fe$_{1+x}$Te, the FM spin stripes are
doubled and rotated by 45 degree with respect to the non--magnetic unit cell
(smallest dotted squares). According to the neutron scattering experiment by
Li \textit{et al.} \cite{Li}, this doubled stripe phase realizes at the
smallest possible $x\sim 0.068$, \textit{i.e.\ }close to the stoichiometric
FeTe ($x=0$). The $J_{1a}$, $J_{1b}$, $J_{2a}$, and $J_{2b}$ represent the
first nearest neighbor AFM, FM, second nearest AFM, and FM interactions,
respectively. While the second neighbor coupling is always AFM in arsenides
(Fig.~\ref{schem}(a)), both FM and AFM second neighbor couplings exist in
FeTe (Fig.~\ref{schem}(b)). As $x$ increases further, the spin ordering
becomes incommensurate at $x\sim 0.141$ and the incommensurate ordering
vector depends on $x$. Importantly this different magnetic structure found
in Fe$_{1+x}$Te cannot be matched with the ($\pi ,\pi $) nesting which is
common for AFM parent materials of all the arsenide superconductors. The
different spin structure found in Fe$_{1+x}$Te remains as a puzzle because
DFT calculation predicts the same Fermi surface topology and the same ($\pi
,\pi $) nesting \cite{Subedi}.

\begin{figure}[t]
\centering \includegraphics[width=10cm]{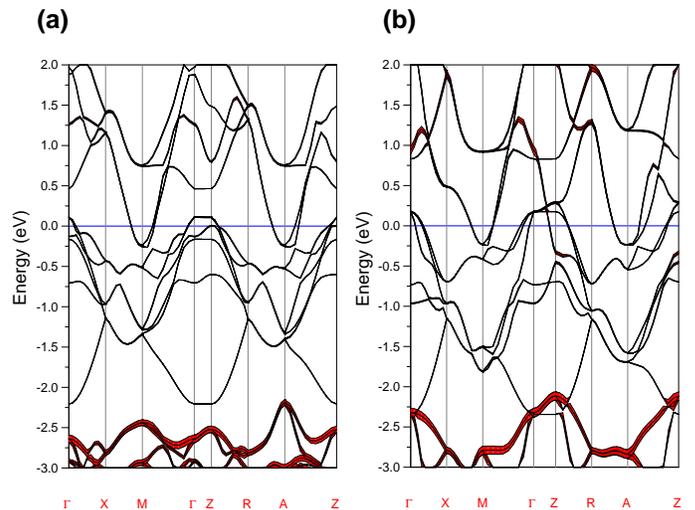}
\caption{(Color online) The calculated band structure for the non-magnetic
phase of (a) LaFeAsO and (b) FeTe. The As-$4p$ and Te-$5p$ bands are
depicted by `fat' bands. Fermi level is set to be zero (horizontal line). }
\label{band}
\end{figure}

Fig.~\ref{band} shows the calculated band dispersions of a typical arsenide
material, LaFeAsO (Fig.~\ref{band}(a)), and FeTe (Fig.~\ref{band}(b)). As
discussed by Subedi \textit{et al.}~\cite{Subedi}, the two band structures
are similar. The Fe--$3d$ states are dominant around the Fermi level while
the anion $p$ bands depicted by `fat' bands are located fairly well below
the Fermi level, and the similar band structure around the Fermi level
produces the same Fermi surface topology. As a result, the same kind of ($%
\pi ,\pi $) nesting is obtained for FeTe as in the other arsenide materials 
\cite{Subedi}. Here we focus on the differences found in the electronic
structure. Firstly, Te--$p$ bands hybridize with Fe--$d$ at around $-2.5$ eV
along $\Gamma $--$Z$ line while As--$4p$ is well separated from Fe--$3d$
states. Different features along $\Gamma $--$Z$ are also found at energies
above the Fermi level. In FeTe, there are significant band crossings at
about $+0.3$ eV. Another notable difference exists at around the $X$ point
above the Fermi level where the parabolic band along $X$--$M$--$\Gamma $ is
flattening at about +0.5 eV in FeTe. It is also noted that there is no band
crossings across the $X$ point in the range of $+0.5$--$+1.8$ eV. These
differences above the Fermi level suggest that possibly different Fermi
surface topology is induced by electron doping as excess Fe atoms appear in
stoichiometric FeTe host.

\begin{figure}[t]
\centering \includegraphics[width=10cm]{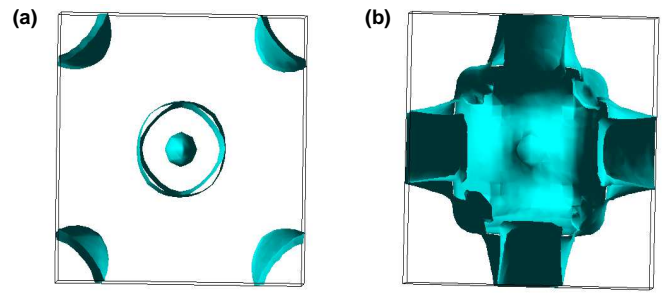}
\caption{(Color online) Fermi surface (in a-b plane) for FeTe as a function
of doping level, $\protect\delta$: (a) $\protect\delta \sim$ 0.0, (b) $%
\protect\delta \sim 0.5$ electron per formula unit. The corner and center of
the square unitcell corresponds to the $\Gamma$ and $M$ point, respectively. 
}
\label{frs}
\end{figure}

Fig.~\ref{frs} shows the Fermi surface for Fe$_{1+x}$Te as a function of
doping: (a) without doping and (b) doping by $0.5$ electrons per formula
unit. Fig.~\ref{frs}(a) is in good agreement with previous calculation \cite%
{Subedi} and clearly shows the existence of ($\pi ,\pi $) nesting as in the
arsenides. The most important feature found in Fig.~\ref{frs}(b) is the
square--type topology developed at around $\Gamma $ and ($\pi ,0$) points
with a similar size. It suggests a new nesting at ($\pi ,0$), which is
consistent with the rotated and doubled stripe spin structure (Fig.~\ref%
{schem}(b)). The doping level by $\delta \sim 0.5$ electrons would
correspond to Fe$_{1.063}$Te provided all eight valence electrons of excess
Fe atoms contribute to the change in the Fermi level within a simple rigid
band approximation. According to a recent neutron scattering by Li \textit{%
et al.}, the commensurate doubled stripe (Fig.~\ref{schem}(b)) spin
structure is realized \cite{Li} in Fe$_{1.068}$Te which is very close to our
case. Such agreement assumes that our simplified rigid band treatment of
doping may indeed capture the essential physics of this system.

\begin{figure}[t]
\centering \includegraphics[width=9cm]{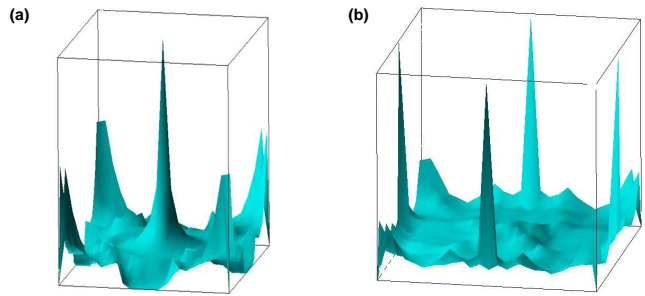}
\caption{(Color online) The imaginary part of calculated Stoner
susceptibility, $\protect\chi_0 (\mathbf{q}_z=0)$, for FeTe in arbitrary
unit: (a) $\protect\delta \sim$ 0.0, (b) $\protect\delta \sim 0.5$ electron
per formula unit. }
\label{hiq}
\end{figure}

The nesting property is further examined by calculating the Stoner response
function, $\chi _{0}$, given by 
\begin{equation}
\chi _{0}=\frac{f(\epsilon _{\mathbf{k}})-f(\epsilon _{\mathbf{k+q}})}{%
\epsilon (\mathbf{k})-\epsilon (\mathbf{k+q})-\omega -i\delta }.
\end{equation}%
Fig.~\ref{hiq}(a) shows the imaginary part of $\chi _{0}(\mathbf{q}_{z}=0)$
for the stoichiometric FeTe without doping (no excess Fe) in which the ($\pi
,\pi $) nesting is clearly seen and is consistent with its Fermi surface in
Fig.~\ref{frs}(a). The \textit{Im}{$\chi _{0}(\mathbf{q}_{z}=0)$} for $%
\delta =0.5$ (equivalent to Fe$_{1.063}$Te) is shown in Fig.~\ref{hiq}(b).
Note the strong ($\pi ,0$) nesting and the suppressed ($\pi ,\pi $) nesting,
which demonstrates the remarkable difference of the doped FeTe from the
undoped FeTe and Fe arsednies. The novel ($\pi ,0$) nesting is derived from
the square--type Fermi surfaces shown in Fig.~\ref{frs}(b) and is driven by
the electron doping through the excess Fe atoms. It is noted that the small
amount of excess Fe plays the key role in determining the magnetic structure
as the result of the Fermi surface change. Since this ($\pi ,0$) nesting
matches with the rotated--doubled spin stripe found in experiment, our
result strongly supports the same spin fluctuation mechanism for
superconductivity in Fe chalcogenides as in the Fe arsenides.

The nesting property of Fe$_{1.063}$Te at $\mathbf{q}_{z}\neq 0$, is
different from that at $\mathbf{q}_{z}=0$. While any notable peak is not
found in \textit{Im}$\chi (\mathbf{q}_{z}\neq 0)$ for the stoichiometric
undoped FeTe, the \textit{Im}{$\chi _{0}(\mathbf{q}_{z}\neq 0)$} of Fe$%
_{1.063}$Te shows rather complicated features including both ($\pi ,\pi $)
and ($\pi ,0$) peaks with reduced intensities. These features are attributed
to a significant variation of the Fermi surface along the $Z$--direction
which is also reflected in the band dispersion in Fig.~\ref{band}. The
doping level of $\delta \sim 0.5$ corresponds to the Fermi level shifted by
about 0.76 eV where it can be noted that the FeTe bands along $\Gamma $--$Z$
direction around $\sim +0.76$ eV region are different from those of LaFeAsO
(Fig.~\ref{band}).

By further dopings, the Fermi surface topology once again changes
significantly, and the ($\pi ,0$) nesting disappears. At the doping level of 
$\delta \sim 1.1$, which would correspond in our analysis to Fe$_{1.141}$Te
with the incommensurate spin order as observed in the experiment \cite{Li},
the ($\pi ,0$) nesting is largely suppressed. At this doping level,
square--type structures are no longer found in the Fermi surface, and the
complicated multiple--peak structure is found to be developed in the $\chi
_{0}$ plot. It might be responsible for the experimentally observed
incommensurate spin orderings \cite{FeSeTe-Incomm,Li}. While, at $\mathbf{q}%
_{z}=0$, neither ($\pi ,\pi $) nor ($\pi ,0$) peak is found in $\chi _{0}$,
several prominent peaks exist at around $\Gamma $ point which are gradually
suppressed along the ($\pi ,\pi $) line. At $\mathbf{q}_{z}=2\pi /3c$,
notable four peaks are found around the ($\pi ,\pi $) points. This structure
found at $\mathbf{q}_{z}\neq 0$ shows another difference of Fe$_{1+x}$Te
from FeAs materials and is originated from the different band structures and
hybridizations around $Z$ point above the Fermi level as shown in Fig.~\ref%
{band}. The accounting for the excess Fe atoms and the validity of the rigid
band approximation is a highly non--trivial problem even when we tried to
simulate the doping by supercell calculations. For the present study,
however, our excellent agreement with experiment demonstrates that our
simplified treatment well describes this material.

The magnetic properties are summarized in Table\ \ref{tab}. Note that the
calculated magnetic moment based on the experimental $z(Te)$ is about 2.09 $%
\mu _{B}$ which is in good agreement with the neutron data $\sim 1.97\mu _{B}
$. This result demonstrates another difference between FeTe and Fe
arsenides. In the arsenides, it is known that using experimental $z(As)$ in
the calculation always leads to overestimating the Fe moment \cite%
{Yin,Mazin-PRB,Han2}. For comparison, we also present in Table~\ref{tab} the
data for parent material, LaFeAsO. The numbers in parenthesis are the
experimental ones. The overestimation by LSDA is more than a factor of four,
which is exceptionally large. The origin of this large discrepancy is still
under debate \cite{Mazin-PRB}. If the small moment observed in experiment is
attributed to domain motions as suggested by Mazin and Johannes \cite%
{Mazin-NatPhys}, the good agreement found in FeTe implies that this system
is free from such dynamics, which calls the further investigation along this
line.

Table~\ref{tab} also shows the calculated exchange couplings of the
Heisenberg spin hamiltonian, $H=J\sum_{<i,j>}\mathbf{S}_{i}\cdot \mathbf{S}%
_{j}$, in FeTe compared to the arsenides. Importantly, without doping, even
on top of the double--stripe spin ordering, our linear response calculations
predict the unstable magnetic interactions. While the charge density
self--consistency is achieved for both single-- and double--stripe spin
order, the spin waves for the undoped FeTe are found to be unstable with
respect to the spin angle tilting. For the single--striped FeTe, $J_{1b}$
becomes AFM and its strength is much larger than $J_{2a}$ which is in a
sharp disagreement with the single--stripe--stabilized LaFeAsO. In the
undoped double stripe FeTe, the overall size of exchange interactions is
small, and the signs of $J_{1a},J_{1b},J_{2a}$ do not correspond to the
actual ordering. It implies that the stoichiometric FeTe ($x=0.0$) is hardly
stabilized, which partly explains the reason that the FeTe sample always has
some amounts of excess Fe atoms \cite{Li}.

The spin waves and magnetic interactions are stabilized in Fe$_{1.063}$Te
being consistent with experiment. In arsenides, the exchange interactions
are represented by two major AFM interactions, $J_{1a}$ and $J_{2}$, and
their strengths are in the same range as is seen in the entire arsenide
family \cite{Han2}: The $J_{1a}\sim 45$ meV and $J_{2}\sim 20$ meV while the
ferromagnetic $J_{1b}$ is very small. Based on these exchange interactions,
the striped AFM phase is stabilized. In Fe$_{1.063}$Te, on the other hand, $%
J_{2a}$ is the strongest while $J_{1a}$ is as small as $J_{1b}$ in the
arsenides. It is noted that the two FM coupling, $J_{1b}$ and $J_{2b}$, are
fairly large which might be responsible for the novel doubled stripe AFM
spin order. Our results suggest the spin wave velocities and dispersions for
Fe chalcogenides are different from the arsenides, which can be verified by
inelastic neutron scattering.

\begin{table}[tbp]
\begin{tabular}{ccccccc}
\hline
& System & Moment & $J_{1a}$ & $J_{1b}$ & $J_{2a}$ & $J_{2b}$ \\ \hline
double stripe & Fe$_{1.068}$Te & 2.09 (1.97\footnotemark[2]) & $-$7.6 & $-$%
26.5 & 46.5 & $-$34.9 \\ 
($\pi,0$) & FeTe & 2.16 & $-$4.2 & 12.9 & $-$6.2 & $-$15.3 \\ \hline
single stripe & FeTe & 2.09 & 38.6 & 21.7 & 5.0 & -- \\ 
($\pi,\pi$) & LaFeAsO\footnotemark[1] & 1.69 (0.36\footnotemark[3]) & 47.4 & 
$-$6.9 & 22.4 & -- \\ \hline
\end{tabular}
\footnotetext[1]{Ref.~\cite{Han2}} 
\footnotetext[2]{Ref.~\cite{Li}} 
\footnotetext[3]{Ref.~\cite{Cruz}}
\caption{ The calculated Fe moment (in $\protect\mu_B$) and exchange
parameters (in meV) for double stripe Fe$_{1.068}$Te (doped) and FeTe
(undoped) along with the single stripe FeTe (undoped). The results of
LaFeAsO are also presented for comparison and the experimental moments
are given in parenthesis.}
\label{tab}
\end{table}

In conclusion, our electronic structure calculations show that a small
amount of excess Fe atoms existing in the Fe$_{1+x}$Te samples changes the
Fermi surface topology significantly. As a result, a novel ($\pi ,0$)
nesting appears at $x\approx 0.063$, and the rotated double--stripe AFM spin
structure stabilizes. This is different from Fe arsenide parent materials.
Our finding of the `missing nesting' explains the origin of the spin density
wave observed by a recent neutron experiment and validates the spin
fluctuation theory of superconductivity for Fe chalcogenides. Further doping
suppresses the ($\pi ,0$) nesting and produces a multi--peak structure in
the Stoner susceptibility which might be responsible for the incommensurate
spin order observed in experiments at higher levels of excess Fe atoms. The
calculated exchange interactions and spin moment demonstrate the role of
excess Fe atoms in stabilizing the magnetic structure, and imply a different
magnetic behavior of chalcogenide superconductors from the arsenides.

This work is supported by NSF Grant DMR--0606498, and DOE SciDAC Grant
SE--FC0206ER25793.

\end{document}